\documentclass[aip, jcp, reprint, floatfix, a4paper]{revtex4-1}
\pdfoutput=1
\usepackage{amsmath, amssymb}
\usepackage{graphicx}
\usepackage[caption=false]{subfig}
\usepackage{bm}
\usepackage{siunitx}
\usepackage{braket}
\usepackage{placeins}
\graphicspath{{./images/}}

\begin{document}

\title{Optimization of energy transport in the Fenna-Matthews-Olson complex via
       site-varying pigment-protein interactions}

\author{S. A. Oh}
\affiliation{Dodd-Walls Centre for Photonic and Quantum Technologies and
             Department of Physics, University of Otago, Dunedin 9016,
             New Zealand}
\author{D. F. Coker}
\email{coker@bu.edu}
\affiliation{Department of Chemistry, Boston University, Massachusetts 02215,
             United States}
\author{D. A. W. Hutchinson}
\email{david.hutchinson@otago.ac.nz}
\affiliation{Dodd-Walls Centre for Photonic and Quantum Technologies and
             Department of Physics, University of Otago, Dunedin 9016,
             New Zealand}
\affiliation{Centre for Quantum Technologies, National University of Singapore,
             3 Science Drive 2, Singapore 117543, Singapore}

\date{\today}

\begin{abstract}
    Energy transport in photosynthetic systems can be tremendously efficient. In
    particular we study exciton transport in the Fenna-Mathews-Olsen (FMO)
    complex found in green sulphur bacteria. The exciton dynamics and energy
    transfer efficiency is dependent upon the interaction with the system
    environment. Based upon realistic, site-dependent, models of the system-bath
    coupling, we show that this interaction is highly optimised in the case of
    FMO. Furthermore we identify two transport pathways and note that one is
    dominated by coherent dynamics and the other by classical energy
    dissipation. In particular we note a strong correlation between energy
    transport efficiency and coherence for exciton transfer from
    bacteriochlorophyll (BChl) 8 to BChl  4. The existence of two clear pathways
    and the role played by BChl 4 also challenges assumptions around the
    coupling of the FMO complex to the reaction centre.
\end{abstract}


\maketitle

\section{\label{sec:Introduction}Introduction}
Given the importance of the effect of the environment upon energy transport in
photosynthetic systems, it is common practice in theoretical studies to assign
to each chromophore, identical spectral densities, usually described by a
smooth, simple functional form. While this simplistic approach should still
allow the essential physics to be captured, it may nevertheless mask the
influence of realistic site variations which could potentially contain
interesting physics in itself.

In fact, recent computational chemistry calculations have shown, for instance in
the case of the Fenna-Matthews-Olson (FMO) complex of the green sulfur bacteria,
that site variations of spectral density can be significant
\cite{lee2016modeling, rivera2013influence, olbrich2011theory}. Obviously, in
biological systems, traits showing some degree of heterogeneity are expected,
and in many cases, it is simply random. However, the green sulfur bacteria, due
to its ability to survive in extremely low-light conditions, must certainly
possess highly-evolved light-harvesting machinery. It is then only natural to
suspect that there may be an underlying design principle associated with this
heterogeneous system-bath couplings. A previous publication demonstrated that
the heterogeneous spectral densities can significantly impact exciton transport
dynamics and suggested that this could be a mechanism used to tune energy
transport\cite{rivera2013influence}.

In addition, exciton delocalization may also be affected by site-dependent
system-bath couplings. Sato and Reynolds \cite{sato2014resonant} found that the
length and robustness of quantum coherence in a generic dimer is simultaneously
increased under specific ratios of site energy mismatch to site reorganization
energy mismatch. Due to the presence of long-lived coherence within the
timescale of exciton dynamics and the ability of quantum walks to sample
multiple paths simultaneously, there is a large amount of speculation that
quantum coherence may remain important to the remarkable efficiency of
photosynthetic energy transport. It is generally accepted that the dynamics are
modified by coherence and that optimal transport occurs in the intermediate
regime between the coherent and incoherent limits\cite{mohseni2008environment,
plenio2008dephasing, ishizaki2009theoretical, rebentrost2009environment,
caruso2009highly, chin2010noise}. However, it remains an open question whether
or not, in this optimal regime, comparable or even better efficiencies can be
achieved by a more classical type of mechanism, e.g. energy funneling. As such,
it is entirely plausible that the presence of quantum coherence is to serve a
different biological function or that it is simply a by-product of dense
chromophore packing, with no specific function of its own. Finding evidence of
underlying coherence optimization and a system design which can effectively
utilize coherence would certainly provide some confidence regarding its
importance in energy transport.

In this work, we present evidence that the site-varying spectral densities of
the FMO complex are indeed optimized for energy transport. This finding has been
made possible by the availability of accurate site-dependent spectral densities
from recent quantum chemistry/molecular dynamics calculations
\cite{lee2016modeling, lee2016semiclassical}.

The basic theories of energy transport or spectroscopic response assume that the
full Hamiltonian can be broken into a system part, $\hat{H}_S$, a bath, or
environmental component, $\hat{H}_B$, and an interaction between these system
and bath subsystems, $\hat{V}$. The partitioning of the different degrees of
freedom between the system and bath subsystems is determined by the resolution
capabilities of a given experiment. Thus, for example, high resolution 2D
electronic spectroscopy (2DES) can actually resolve transitions between higher
frequency intramolecular vibrational and electronic states, so in such
experiments the system part is best described by the vibronic eigenstates of the
chromophores and the bath becomes the lower frequency environmental, or
intermolecular degrees of freedom. For less well resolved experiments, for
example, monitoring energy migration from one chromophore to another and not
resolving (i.e. averaging over) the intramolecular vibrational structure of the
chromophores, the system is best described by their coupled electronic states,
while the bath becomes all the intra- and intermolecular vibrations. The study
presented here approximately adopts this latter perspective in that it assumes
that $\hat{H}_S$ is represented by the eigenstates of the coupled electronic
subsystem, while the intermolecular spectral density composed of a continuum of
protein environmental modes describes the bath. In these studies we will
disregard the influence of the discrete intramolecular component of the spectral
density as they are expected to play a fairly minor role in energy transport in
chlorophyll systems with small intramolecular Huang-Rhys factors. The
possibility of further optimization from vibronic contribution and interplay
with the intermolecular component of the spectral density will be explored in a
future publication. Preliminary results from these further studies do not
materially affect the conclusions in this paper.

\section{Theoretical model}
The total system-bath Hamiltonian is expressed as
\begin{eqnarray}
    \hat{H}_{tot} = \hat{H}_s + \hat{H}_b + \hat{H}_{sb},
\end{eqnarray}
where $\hat{H}_s$, $\hat{H}_b$ and $\hat{H}_{sb}$ are the system (electronic),
bath and system-bath interaction Hamiltonians respectively.

\subsection{The electronic system}

The electronic system consists of the chromophores in the pigment-protein
complex (PPC). Its Hamiltonian governs the coherent part of the evolution and is
described by the tight-binding model in the site basis $\ket{m}$:
\begin{eqnarray}
    \hat{H}_s = \sum_{m=1}^{N} E_m\ket{m}\bra{m} + \sum_{m\neq n}^{N} V_{mn}\ket{m}\bra{n},
\end{eqnarray}
where $E_m$, $V_{mn}$ and $N$ are the site energies, electronic coupling between
pigments $m$ and $n$ and number of chromophores respectively. Since our system
operates in a low-photon environment, the single exciton manifold approximation
is valid here.  The $k$th eigenstate of the system Hamiltonian (also known as
the exciton state) can be decomposed in terms of the site basis:
\begin{eqnarray} \label{eq:M_superposition_of_m}
    \ket{k} = \sum\limits_{m} c_{m,k}\ket{m}.
\end{eqnarray}
In this work, we use the 8-site Hamiltonian for the \textit{Prosthecochloris
aestuarii} (\textit{P. aestuarii}) species as presented by Moix et. al
\cite{moix2011efficient}.

\subsection{The bath and system-bath coupling}
The protein environment is commonly modelled as a bath of harmonic oscillators with $\hat{H}_b = \sum\limits_{i,m}\Big(\hbar \omega_{i,m} b_{i,m}^{\dagger} b_{i,m} + \frac{1}{2}\Big)_m$ where $b_{i,m}^{\dagger}$($b_{i,m}$) are the creation (annihilation) operators of excitations of the $i$th bath mode with frequency $\omega_{i,m}$ on pigment $m$. In this work, the phonon modes on each site are treated as being uncorrelated with each other and coupled linearly to the diagonal part of the system Hamiltonian such that:
\begin{eqnarray}
\hat{H}_{sb} = \sum\limits_{i,m} u_{i,m}(b_{i,m}^{\dagger} + b_{i,m})\ket{m}\bra{m}.
\end{eqnarray}
Here $u_{i,m}$ is the coupling between the electronic transition of the $m$th site and the $i$th phonon mode. Physically, this equation reflects the effect of the protein environment dynamically modulating the site energies of the pigments. All information about the system-bath interaction of each pigment $m$ is contained in its corresponding spectral density $J_m(\omega) = \pi\sum\limits_{i}|u_{i,m}|^2\delta(\omega - \omega_{i,m})$. The reorganization energy $\lambda_m$ of pigment $m$ is in turn related to $J_m(\omega)$ by the following integration over frequency $\omega$:
\begin{eqnarray}
\lambda_m = \dfrac{1}{\pi}\int_0^{\infty} \dfrac{J_m(\omega)}{\omega} d \omega.
\label{eq:reorganization_energy}
\end{eqnarray}

\section{\label{sec:Methods}Numerical methods}
The numerical computation of the dynamics was performed using the Modified Redfield Theory (MRT) \cite{zhang1998exciton} and its more recent upgrade, the Coherent Modified Redfield Theory (CMRT)\cite{hwang2015coherent,chang2015accuracy,ai2014efficient,tao2016proposal}. It has been shown that MRT and CMRT are reasonably valid over a broad range of system-bath coupling strengths and provide reasonable agreement with dynamics computed from numerically exact methods \cite{yang2002influence,hwang2015coherent,ai2014efficient}. This, coupled with the convenience and efficiency as compared to numerically exact methods make them suitable numerical methods in this work.

In contrast to MRT, CMRT allows the computation of coherence terms in the density matrix as well as the incorporation of some non-Markovianity. Also while the MRT population vector $P(t)$ can only be in the exciton basis, the CMRT density matrix $\rho(t)$ can be obtained in both the site and exciton basis, with the site basis being the default (Appendix \ref{app:nmqj}). The downside of CMRT however, and the reason why both versions of the same method had to be used, is its much longer computation time. This renders it infeasible for the statistical methods and genetic algorithm used in Sections \ref{sec:histogram} and \ref{sec:gen_algo} respectively, which incidentally require only the population terms. We thus employ MRT in those cases while using CMRT for the rest of the work in this paper. As in Refs. \onlinecite{ai2014efficient,tao2016proposal}, we solve for the non-Markovian dynamics of CMRT with the non-Markovian Quantum Jump (NMQJ) technique \cite{piilo2008non,piilo2009open}.

To justify the use of MRT in place of CMRT, we have checked that the effects of non-Markovianity are not overly significant, or in other words, the general trend of population dynamics (in the exciton basis) is similar with the two versions. For self-consistency, we give a brief outline of MRT, CMRT and NMQJ in Appendices \ref{app:MRT}, \ref{app:CMRT} and \ref{app:nmqj} respectively. The full derivation and additional details can be found in Refs. \onlinecite{zhang1998exciton,yang2002influence} for MRT, Refs. \onlinecite{hwang2015coherent, chang2015accuracy, ai2014efficient, tao2016proposal} for CMRT and Refs. \onlinecite{piilo2008non, piilo2009open} for NMQJ.  All computation is performed for the physiological temperature of $T = 300 K$.

\section{\label{sec:Setup}Numerical setup}
The two most important pigments in the FMO are the initial excitation site and the target site –-- the former influences the exciton dynamics and the latter determines our assessment of energy transport efficiency.
It is now believed that BChl 8 is the most likely linker site between the chlorosome and the rest of the FMO complex \cite{schmidt2010eighth}. However, BChl 8 is normally lost during sample preparation, and before its recent discovery, \cite{ben2004evolution,tronrud2009structural} BChls 1 and 6 were proposed as the possible linker sites. These two pigments are also the usual initial photoexcitation sites in spectroscopic experiments. For these reasons, many theoretical studies of the past and present employ BChls 1 and 6 as the initial excitation site. Since our motivation is to understand the in vivo workings of the FMO, and not for comparison to spectroscopic data, we choose BChl 8 for this work.

In the literature, the linker site to the reaction centre is usually assumed to be BChl 3, which is also the lowest energy pigment with the closest proximity to the reaction centre. There exists some ambiguity, however. Wen et. al. \cite{wen2009membrane} reported that it is the BChl 3 side of the FMO complex which interacts with the reaction centre, but their experiment did not pinpoint the exact pigment(s). Furthermore, it has been mentioned in several publications \cite{adolphs2006proteins,ishizaki2009theoretical,ritschel2011absence,zhu2011modified,cheng2009dynamics,shabani2014numerical} that the two lowest energy pigments, BChls 3 and 4, are in the target region close to or in contact with the reaction centre. As such, we consider both BChls 3 and 4 as the possible target sites in this paper.

To aid physical interpretation, we represent the realistic spectral densities in the Drude-Lorentz regularized Ohmic form \cite{mukamel1999principles}:
\begin{eqnarray}
J_m^{DL}(\omega) = 2 \lambda_m \Omega_m \dfrac{ \omega}{\omega^2 + \Omega_m^2},
\label{eq:drude_lorentz}
\end{eqnarray}
where $\lambda_m$ and $\Omega_m$ are the reorganization energy and cutoff frequency of site $m$ respectively. Here the values of $\lambda_m$ are calculated from the realistic site spectral densities via eq \ref{eq:reorganization_energy}. $\Omega_m$, which corresponds to the inverse of the bath correlation time, can be obtained by first computing the site-dependent bath correlation function (eq. \ref{eq:bath_corr_function}). For simplicity, we assume that the real part of the bath correlation functions can be represented by time decaying exponential functions, from which it is possible to extract the bath correlation time and subsequently $\Omega_m$. We have verified that the dynamics with the realistic site-dependent spectral densities can be reliably reproduced with their Drude-Lorentz forms.

The computed values of $\lambda_m$ and $\Omega_m$ are presented in Table \ref{tab:reorg_cutoff_freq}. It is clear that the FMO complex has a fairly significant range of $\lambda_m$ values where the largest is more than 2.5 times the magnitude of the smallest. It is interesting to note that the mean reorganization energy is also in a sense the most representative since half of the pigments have $\lambda_m$ values close to this value.

\begin{table}
\caption{\label{tab:reorg_cutoff_freq} Site-dependent reorganization energies $\lambda_m$ and cutoff frequencies $\Omega_m$ of the FMO Drude-Lorentz spectral density used in this work.}
\begin{ruledtabular}
\begin{tabular}{lccccccc}
\mbox{m}&$\lambda_m$ (\si{cm^{-1}})&$\Omega_m$ (\si{cm^{-1}})  \\
\hline
1 &21.28 & 40.96 \\
2 &31.52 & 88.04 \\
3 &22.86 & 43.52 \\
4 &17.88 & 48.79 \\
5 &15.36 & 52.10 \\
6 &23.18 & 43.55 \\
7 &24.89 & 39.35 \\
8 &41.00 & 37.31 \\
\hline \\
Mean &24.75 &49.20
\end{tabular}
\end{ruledtabular}
\end{table}

The complexity of the problem is now significantly reduced since the spectral density is characterized by only two physically meaningful bath parameters, i.e. $\lambda_m$ and $\Omega_m$. It turns out the problem can be further simplified. By comparing the dynamics obtained with site variation in only one of the bath parameters to that obtained with a completely site-independent benchmark, we established that it is primarily the site variation in $\lambda_m$ which modifies the dynamics, with very little contribution from the $\Omega_m$ site variation. Here the site-independent case is constructed by assigning the mean value of $\lambda_m$ and/or $\Omega_m$ from Table \ref{tab:reorg_cutoff_freq}. Importantly, we note an improvement in exciton transport to the target sites with the site-varying spectral density.

Therefore, all computation will be performed using the Drude-Lorentz spectral density representation, keeping the site-dependent $\Omega_m$ fixed to the original configuration values in Table \ref{tab:reorg_cutoff_freq} and varying only $\lambda_m$.

\section{Results}
\subsection{Optimality of system-bath coupling configuration} \label{sec:histogram}
\begin{figure}
\subfloat[]{\includegraphics[scale=0.7]{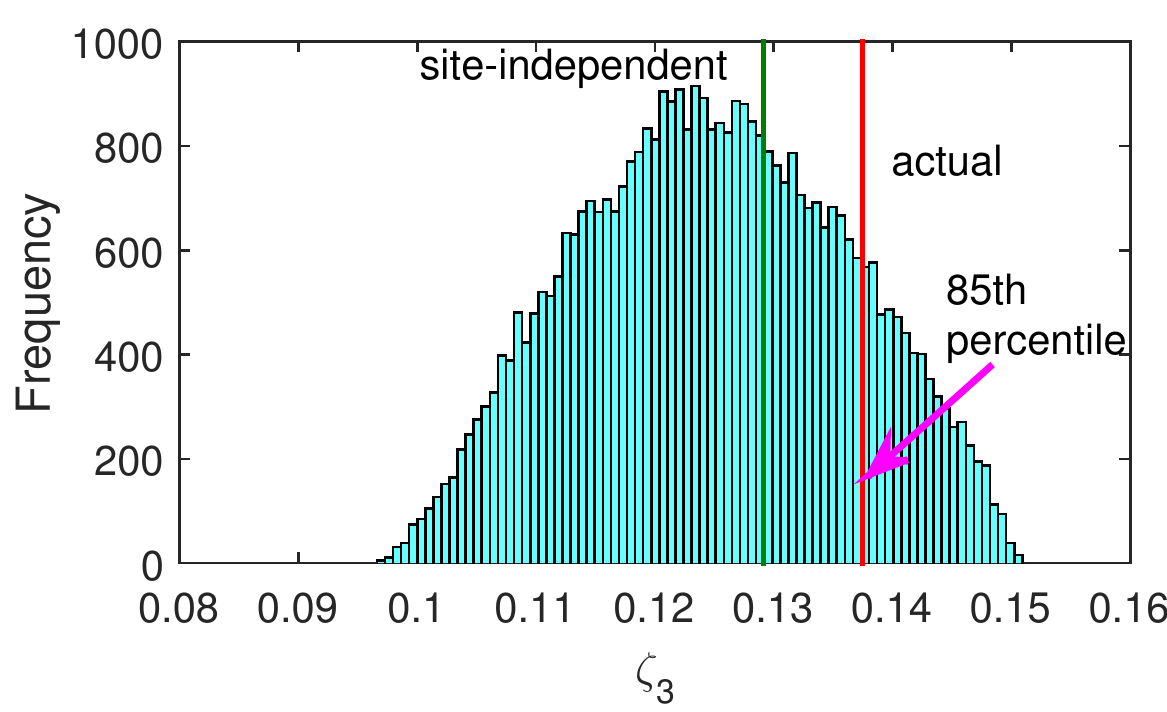}} 

\subfloat[]{\includegraphics[scale=0.7]{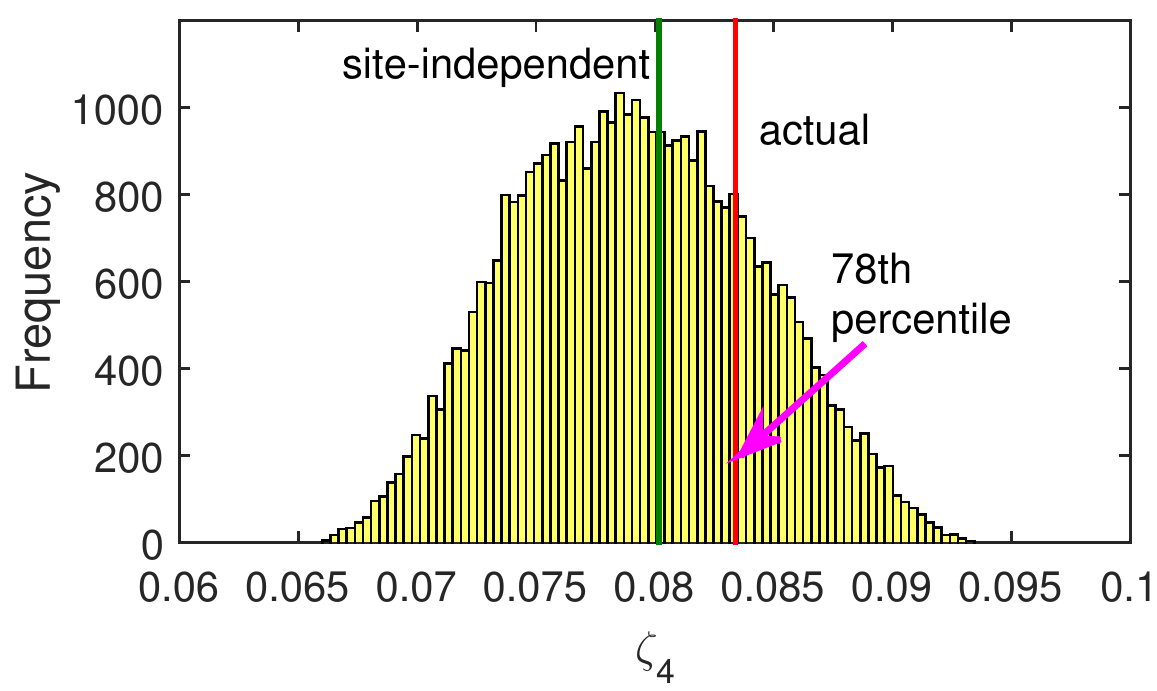}}
\caption{Histogram of efficiencies $\zeta_{m}$ for all possible site permutations of $\lambda_m$ for exciton state (corresponding target site) of (a) k = 1 (BChl 3) and (b) (k = 2) (BChl 4). In both histograms, the original configuration is also included in the sample. The red and green vertical lines indicate the position on the histogram of the original and site-independent configurations respectively. Numerical method: MRT.}
\label{fig:hist_rand_perm}
\end{figure}

Following the observation of energy transport enhancement with the site-dependent spectral density, a more rigorous assessment of optimality is necessary. To this end, the energy transfer efficiency with the FMO site-dependent $\lambda_m$ configuration must be compared to that from a large sample of random site-dependent $\lambda_m$ configurations. Here the energy transfer efficiency, as obtained using the MRT formulation, is defined as the time-averaged population at the predominant exciton state of the target site:
\begin{eqnarray} \label{eq:ETE}
\zeta_{m} = \dfrac{1}{\tau}\int_{0}^{\tau} P_k(t) dt,  \quad \text{where $|c_{m,k}|^2$ is maximal.}
\end{eqnarray}
Here $P_k$ is the population in exciton state $k$, with $k = 1(2)$ for the target site $m = 3(4)$. The participation of an exciton state in each site is denoted by the absolute square of the corresponding expansion coefficient in eq. \ref{eq:M_superposition_of_m} and ranges from 0 to 1. Here $|c_{3,1}|^2 = 0.88$ and $|c_{4,2}|^2 = 0.59$. We have chosen $\tau = 1 \ \si{ps}$ since a large portion of the relevant exciton dynamics and the experimentally observed coherence occur within this timescale.
Two types of randomization were performed. In the first set, the efficiencies of all possible site permutations of the original $\lambda_m$ configuration were computed and presented as a histogram in Figure \ref{fig:hist_rand_perm}.

\begin{figure}
\subfloat[]{\includegraphics[scale=0.75]{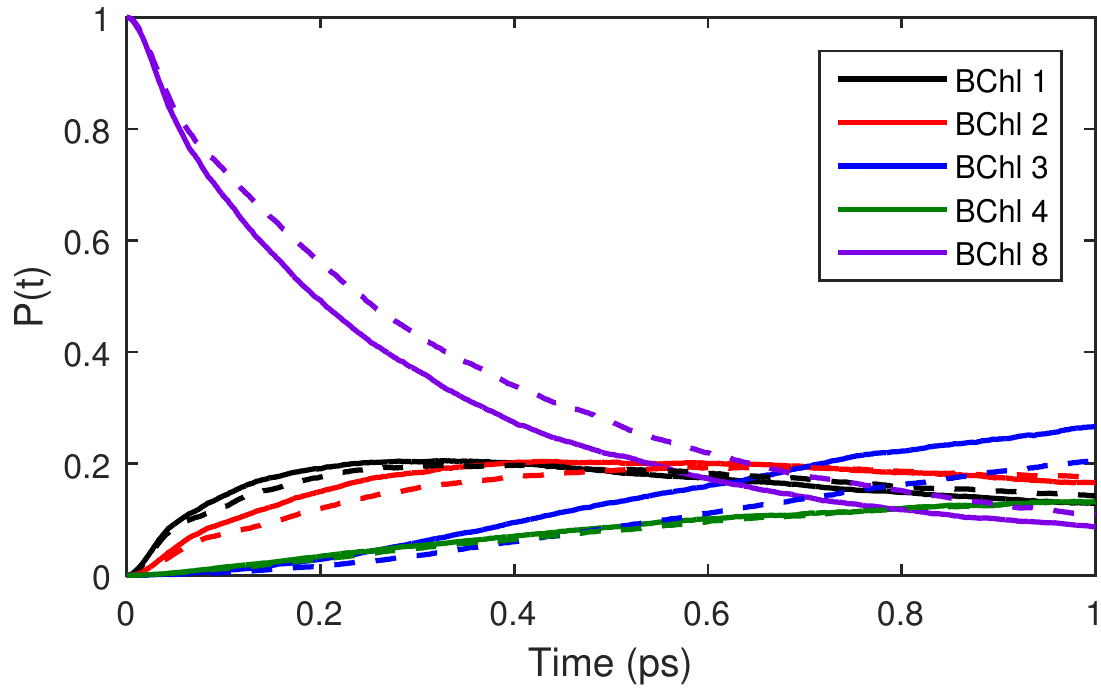}} \\
\subfloat[]{\includegraphics[scale=0.75]{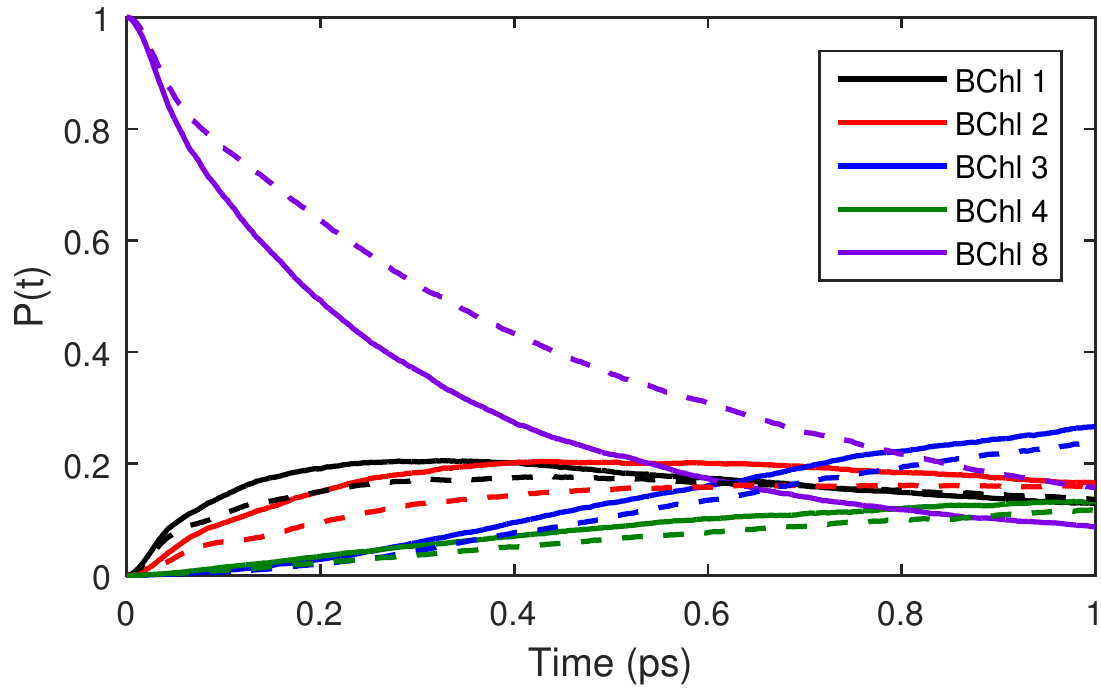}}
\caption{Site population dynamics showing the disparity in energy transport performance with the original FMO $\lambda_m$ configuration (solid curves) and the least efficient site permutation of $\lambda_m$ (dashed curves) for the target site of (a) BChl 3 and (b) BChl 4. Numerical method: CMRT.}
\label{fig:site_dependent_worst_perm3_4}
\end{figure}

Notice that the efficiency of the actual FMO $\lambda_m$ configuration falls in an impressive upper percentile range (85th and 78th percentile for the target sites of BChl 3 and 4 respectively). The fact that it is not the most efficient configuration is not a concern here. In fact, it is not surprising in the context of evolution, since biological constraints may be present and traits only need to be sufficiently functional. In contrast, the site-independent case yields an efficiency closer to the mode of the distribution. A similarly high level of optimization is also observed in our second set of randomization, where now the only constraints on the values are that the mean, maximum and minimum are similar to that of the original configuration. For consistency, the same sample size as the first set, i.e. 40320 was used. As shown in Figure \ref{fig:hist_rand_complete} of Appendix \ref{app:hist_rand_complete}, the efficiency corresponding to the original $\lambda_m$ configuration is in the 87th and 81st percentile for the target sites of BChl 3 and 4 respectively. Once again, the efficiency for the site-independent case appears near the mode of the distribution. To provide a visualization of how a random unoptimized $\lambda_m$ configuration could be detrimental to transport efficiency, we present in Figure \ref{fig:site_dependent_worst_perm3_4} the comparison of the exciton dynamics with the original $\lambda_m$ configuration to that with the least efficient configuration in Figure \ref{fig:hist_rand_perm}.

Finally, it is worth mentioning that when we tested BChls 1 and 6 as the initial excitation sites, no such optimization is present. Compared to the site-independent case, there is no appreciable improvement in energy transport in the case of BChl 1, while for BChl 6, there is in fact a deterioration in efficiency. Assuming the distribution of $\lambda_m$ is a result of evolution to optimize energy transport, this observation is further testament that BChl 8 is likely the first pigment to receive the exciton from the chlorosome.

\subsection{Source of optimality} \label{sec:gen_algo}
Having established that the actual site-dependent $\lambda_m$ configuration is highly optimized for energy transport to both BChls 3 and 4, we seek to identify the underlying mechanism behind its effectiveness. We note that network connectivity and site energy distribution determine how effective a particular mode of energy transport (dissipative or coherent) would be. Moreover, larger reorganization energies are conducive to energy dissipation while smaller reorganization energies are beneficial for the sustenance of quantum coherence. This suggests an optimized interplay between the FMO $\lambda_m$  distribution and the design of the electronic system.

Therefore, to determine if the system design is optimized for a dissipative or coherent mode of transport, we establish whether the best efficiency is obtained with larger or smaller values of $\lambda_m$. This is achieved through the use of a genetic algorithm (Appendix \ref{app:gen_algo}). Here, the fitness function is the same measure of efficiency used in the previous section, i.e. the efficiency at the predominant exciton state as evaluated with the MRT formulation (eq. \ref{eq:ETE}).

Now, for the target site of BChl 3, the optimal configuration is one where all the sites have the maximum $\lambda_m$ value. Clearly, the optimal mechanism of energy transport to this target site must be a dissipative one. On the other hand, for the target site of BChl 4, the optimal values are minimized for the three lowest energy pigments (BChls 2, 3 and 4) and maximized for the remaining pigments. This suggests that a combined coherent and dissipative mechanism leads to optimal energy transport in this case. We note that the minimization for BChls 3 and 4 could partly be due to increased coherence prolonging linear combinations between two sites. This would lead to increased population in BChl 4 on average compared to if the process was dissipative, promoting only downhill energy flow to BChl 3. However, we fail to see any indication of Rabi oscillations suggesting this process is highly overdamped and negligible here.

Next, we performed a second genetic algorithm run, but this time with a constraint on the mean, mimicking the conditions in Section \ref{sec:histogram}. Due to this constraint, it is now possible to assess the significance of each pigment to the energy transport process. This is because the only way now for the algorithm to maximize efficiency is through prioritizing the most important pigments by assigning them the best $\lambda_m$ values, and leaving the remainder to the less influential pigments. Thus, in a dissipative process for example, the more influential a pigment is, the larger the computed optimal $\lambda_m$ would be.

The results from the second genetic algorithm run are tabulated in Table \ref{tab:reorg_ori_vs_ga}, where the computed optimal values are now labelled $\lambda_{m,3}^{ga}$ and $\lambda_{m,4}^{ga}$ for the target sites of BChls 3 and 4 respectively. From the larger values of $\lambda_{m,3}^{ga}$, it is clear that the dominant pathway consists of BChls 8, 1, 2 and 3, while BChls 4, 5, 6 and 7 only have minimal contributions. This, together with the previous finding of a dissipative mode of transport, are consistent with the findings of Moix et. al. \cite{moix2011efficient}. Meanwhile, based on the magnitudes of $\lambda_{m,4}^{ga}$, we can infer that the dissipative part of the process predominantly involves BChls 1, 7 and 8, with less significant contribution from BChls 5 and 6.

\begin{table}
\caption{\label{tab:reorg_ori_vs_ga} Original site-dependent reorganization energies $\lambda_m$ of the FMO complex and the optimal configuration of site-dependent reorganization energies $\lambda_{m,3}^{ga}$ and $\lambda_{m,4}^{ga}$ as determined by a genetic algorithm for exciton states (corresponding target sites) of k = 1 (BChl 3) and (k = 2) (BChl 4) respectively. The mean, lower bound and upper bounds for the solution of the genetic algorithm have been set to be similar to that of the original FMO configuration. Numerical method: MRT.}
\begin{ruledtabular}
\begin{tabular}{lccccccc}
\mbox{m}&$\lambda_m$ (\si{cm^{-1}})&$\lambda_{m,3}^{ga}$ (\si{cm^{-1}})&$\lambda_{m,4}^{ga}$ (\si{cm^{-1}})  \\
\hline
1 &21.28 & 36.37 & 40.95\\
2 &31.52 & 32.44 & 15.37\\
3 &22.86 & 40.52 & 15.36\\
4 &17.88 & 15.36 & 15.36\\
5 &15.36 & 15.36 & 19.95\\
6 &23.18 & 15.36 & 21.27\\
7 &24.89 & 15.36 & 37.79\\
8 &41.00 & 27.21 & 31.92
\end{tabular}
\end{ruledtabular}
\end{table}

The $\lambda_{m,3}^{ga}$ and $\lambda_{m,4}^{ga}$ configurations also provide useful optimality benchmarks. Through comparison with the FMO $\lambda_m$ configuration, the source of its high degree of optimality can be identified. With the exception of $\lambda_1$, all the site reorganization energies show some level of optimization for either one or both of the target sites. The two smallest $\lambda_m$ values, $\lambda_4$ and $\lambda_5$ show a high degree of agreement with the genetic algorithm values for both target sites. $\lambda_2$ and $\lambda_6$ are optimized for only one of the target sites. Meanwhile, the values of $\lambda_3$ and $\lambda_7$ are somewhere in between the optimized values for the two target sites, thus achieving a compromise. Lastly, $\lambda_8$, despite being close to neither of the optimized values, show a similar trend to the genetic algorithm results by virtue of its larger than average value. With regards to the mutual optimization for both target sites, three mechanisms of optimization can be identified. The first is where the magnitude of $\lambda_m$ mutually benefits both target sites, for example the larger than average value of $\lambda_8$ is advantageous for the dissipative type of energy transport to both BChls 3 and 4. The second mechanism, which is also the most interesting, is where the magnitude of $\lambda_m$ benefits only one of the target sites, while minimizing its negative effects on the other target site. This applies when a particular pigment is involved in different effective modes of transport to each of the target sites. A case in point is the small value of $\lambda_4$, which is advantageous for the partially coherent energy transport to the target site of BChl 4. At the same time, it does not overly impede dissipative energy transport to the target site of BChl 3 since the dominant pathway is not involved. Finally, the third mechanism is one where the magnitude of $\lambda_m$ does not assist energy transport to any of the target sites, but the negative impact is simply minimized. For example, the small value of $\lambda_5$ is unfavourable for the dissipative energy transport to both target sites, but the negative effect is mutually minimized since BChl 5 is not on a dominant pathway for any of them.

At this juncture, we must stress that it is not the $\lambda_m$ site variation per se which is responsible for the enhanced performance, since it is really the magnitude of $\lambda_m$ that matters. Rather, it is more likely a case of Nature making the best of an unavoidable situation. Biological constraints in the FMO can lead to site variation in $\lambda_m$; for example, pigments located at the protein-solvent interface tend to have larger reorganization energies than those in the interior \cite{rivera2013influence}. The high efficiency of the FMO $\lambda_m$ configuration compared to various other random configurations of similar average (Section \ref{sec:histogram}) is then simply a consequence of the system and system-bath interaction having evolved to complement each other in a very effective manner, e.g. via selection of appropriate dominant pathways and effective transport mechanisms.

\subsection{Relevance of quantum coherence to efficiency} \label{sec:coherence}

Given the evidence for both dissipative and coherent energy transport pathways, we investigate how energy transport efficiency relates to coherence length in the presence of site-dependent $\lambda_m$. Since the coherence terms of the density matrix are required, all the computation in this section is performed using the CMRT formulation. For the efficiency $\eta_m$, we utilize a measure similar to eq. \ref{eq:ETE} but for the site basis counterpart, i.e:
\begin{eqnarray} \label{eq:ETE_site}
\eta_m = \dfrac{1}{\tau}\int_{0}^{\tau} \rho_{mm}(t) dt,
\end{eqnarray}
where $\rho_{mn}(t)$ are the elements of the time-dependent density matrix computed using the CMRT formulation, with the diagonal elements $\rho_{mm}(t)$ representing the population at site $m$. Here $m = 3(4)$ for the target site of BChl 3(4). As usual, $\tau = 1 \ \si{ps}$.

To quantify the degree of exciton delocalization, we use the coherence length defined by \cite{meier1997multiple,dahlbom2001exciton}
\begin{eqnarray}
L_{\rho}(t) = \dfrac{(\sum_{mn}^{N} |\rho_{mn}(t)|)^2}{N \sum_{mn}^{N} |\rho_{mn}(t)|^2},
\end{eqnarray}
$L_{\rho}(t)$ ranges from 1 for the case of zero coherence to N for a fully delocalized state, i.e. larger $L_{\rho}(t)$ values indicate a larger degree of delocalization. At $t = 0$, even though the state is fully coherent, the population is completely localized at a single site, in which case $L_{\rho}(0) = 1/N$. As a result, $L_{\rho}(t)$ starts from $1/N$ in all cases, increases then peaks after a certain period of time before decreasing. This reflects the scenario of an initially localized pure state becoming more delocalized before the interaction with the environment gradually destroys the coherence. With the original $\lambda_m$ configuration, $L_{\rho}(t)$ peaks at the value of about 2 around $t = 0.7 \ \si{ps}$, corresponding to the fact that within the chosen time scale of $\tau = 1 \ \si{ps}$, the dynamics are still relatively far from thermal equilibrium. Note however, that the state of complete incoherence can never be reached even at equilibrium \cite{moix2012equilibrium}.

In analogy to eqs. \ref{eq:ETE} and \ref{eq:ETE_site}, we define the time-averaged coherence length as:
\begin{eqnarray} \label{eq:L_rho_avg}
L_{\rho,avg} = \dfrac{1}{\tau}\int_{0}^{\tau} L_{\rho}(t) dt.
\end{eqnarray}

\begin{figure}
\includegraphics[scale=0.6]{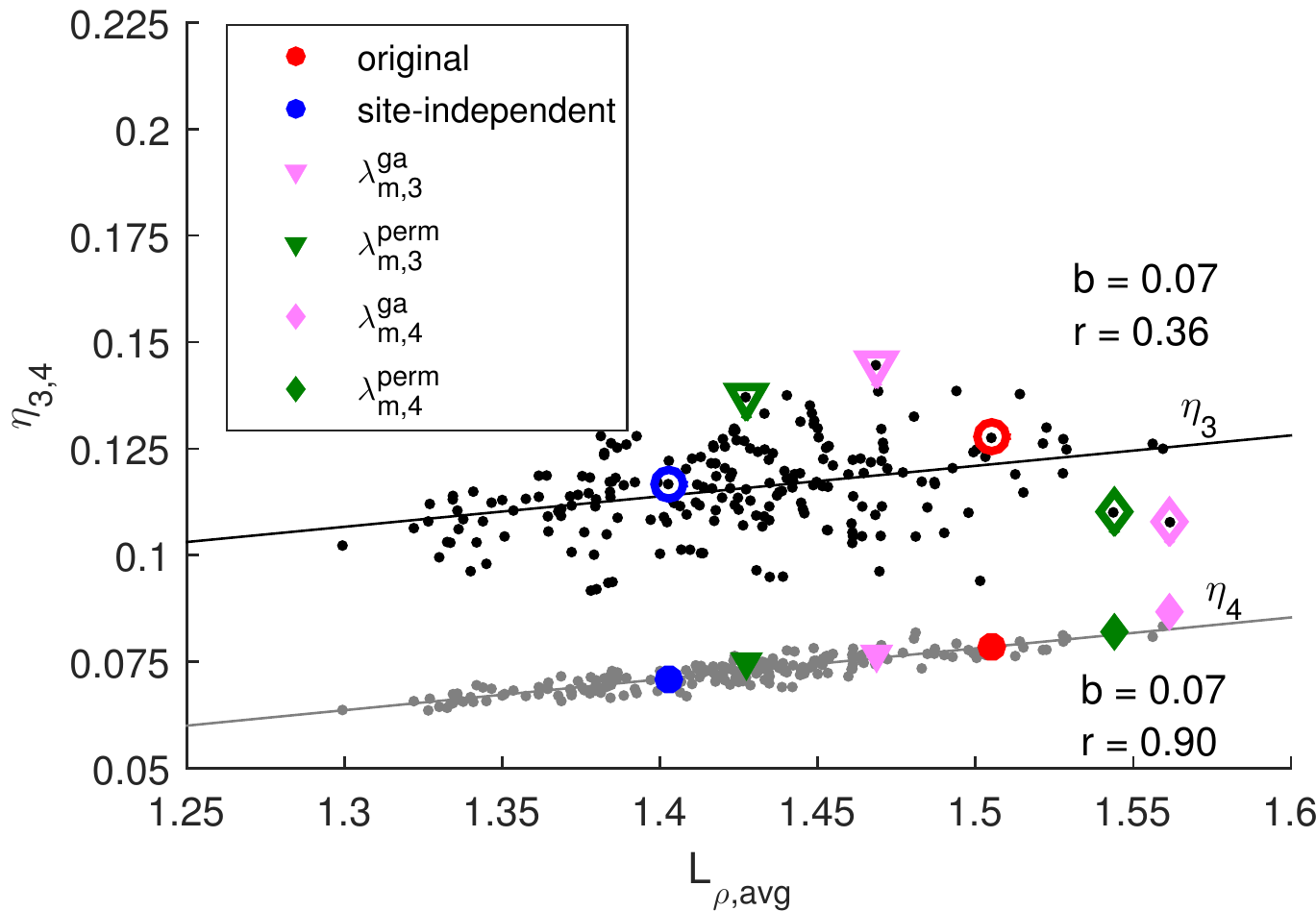}
\caption{Scatter plot of efficiency $\eta_m$ as a function of the time-averaged coherence length $L_{\rho,avg}$ for the assigned target sites of BChls 3 and 4. Black and grey dots represent ($L_{\rho,avg}$,$\eta_3$) and ($L_{\rho,avg}$,$\eta_4$) data points respectively, and correspond to 200 random site-dependent $\lambda_m$ configurations with the same mean, maximum and minimum as the original. Coloured markers represent data points from the following $\lambda_m$ configurations: original, site-independent, optimized $\lambda_m$ configuration from the genetic algorithm (for both target sites BChls 3 and 4) and the most efficient site permutation of original $\lambda_m$ (for both target sites BChls 3 and 4). Open markers are for $\eta_3$ while solid markers are for $\eta_4$. The linear regression line corresponding to $\eta_3$($\eta_4$) is shown in black(grey). $b$ and $r$ are the slope of the regression line and the correlation coefficient respectively. Numerical method: CMRT.}
\label{fig:eta_vs_L_int}
\end{figure}

The relationship between efficiency $\eta_m$ and the time-averaged coherence length $L_{\rho,avg}$ is depicted as a scatter plot in Figure \ref{fig:eta_vs_L_int}. The data points correspond to 200 random site-dependent $\lambda_m$ configuration plus 6 additional relevant data points, namely the original configuration, the site-independent configuration, the optimized solution from the genetic algorithm (i.e. $\lambda_{m,3}^{ga}$ and $\lambda_{m,4}^{ga}$) and the most efficient site permutation of the original configuration (which we shall denote as $\lambda_{m,3}^{perm}$ and $\lambda_{m,4}^{perm}$ for target sites BChls 3 and 4 respectively). For the target site of BChl 3 (Figure \ref{fig:eta_vs_L_int}), even though there is a positive correlation between $L_{\rho,avg}$ and $\eta_m$, it is in the weak to moderate regime, with a correlation coefficient of only 0.36. This relatively weak correlation implies that it is unlikely effects due to coherence are significant in the energy transport mechanism to BChl 3, and further confirms the mainly dissipative nature of the energy transport mechanism as demonstrated in the previous section.

We observe a remarkably strong positive correlation between $\eta_4$ and $L_{\rho,avg}$, with a correlation coefficient of 0.90. We also examined the correlation with $V_{34}$ set to zero in the system Hamiltonian, in order to address the concern that the remarkable correlation could be predominantly attributed to the strong coupling between BChls 3 and 4. While the positive correlation decreased as expected, it remains strong with a correlation coefficient of 0.76. This shows that even though the strong coupling between the two pigments undoubtedly plays a role, it is not overwhelmingly responsible for the correlation between coherence and $\eta_4$.

Figure \ref{fig:eta_vs_L_int} also clearly showcases the capability of the FMO $\lambda_m$ configuration to accommodate efficient energy transport to both target sites, as we have inferred in previous sections. Even though the FMO $\lambda_m$ configuration does not lead to the best efficiency for either of the target sites, it nevertheless corresponds to relatively high efficiency for both target sites. In contrast, two of the most optimized configurations, $\lambda_{m,3}^{ga}$($\lambda_{m,4}^{ga}$) and $\lambda_{m,3}^{perm}$($\lambda_{m,4}^{perm}$) only produce superior efficiencies for their respective target site BChl 3(4), but are significantly less remarkable and are even inferior to the FMO configuration for the other target site BChl 4(3). This illustrates the non-triviality of navigating trade-offs to sufficiently accommodate two largely uncorrelated and partially conflicting processes. The highly coherent nature of $\lambda_{m,4}^{ga}$ and $\lambda_{m,4}^{perm}$ (as reflected by their large $L_{\rho,avg}$ values) while advantageous for the partially coherent energy transport to BChl 4, is unconstructive for the dissipative energy transport to BChl 3 (Figure \ref{fig:eta_vs_L_int}). Similarly, $\lambda_{m,3}^{ga}$ and $\lambda_{m,3}^{perm}$  which only corresponds to moderate coherence only leads to average efficiencies at BChl 4 (Figure \ref{fig:eta_vs_L_int}). Meanwhile, the FMO $\lambda_m$ configuration gives rise to a rather impressive degree of exciton delocalization, where its $L_{\rho,avg}$ value lies between that of $\lambda_{m,3}^{ga}$($\lambda_{m,3}^{perm}$) and $\lambda_{m,4}^{ga}$($\lambda_{m,4}^{perm}$), resulting in intermediate $\eta_3$ an $\eta_4$ values.

\begin{figure}
\includegraphics[scale=0.6]{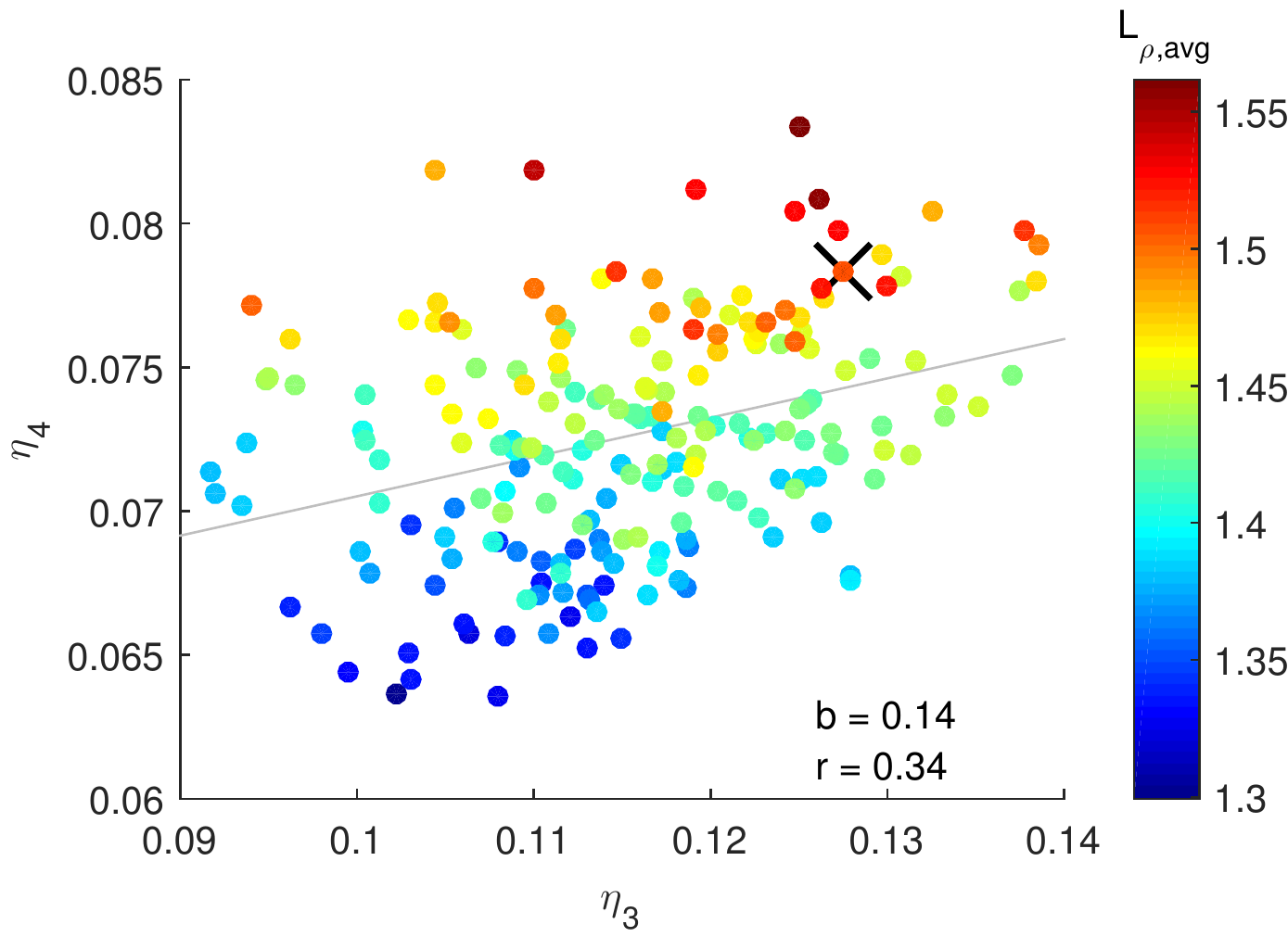}
\caption{\label{fig:eta4_vs_eta3} Scatter plot using the same 206 data points from Figure \ref{fig:eta_vs_L_int} showing the relation between the efficiencies $\eta_3$ and $\eta_4$ at the target sites of BChls 3 and 4 respectively. The black cross indicates the data point corresponding to the original configuration. Numerical method: CMRT.}
\end{figure}

Lastly, we explore the question of independence between the observed optimization of $\eta_3$ and $\eta_4$. This is an important point to address since the relatively strong coupling between BChls 3 and 4 raises the possibility that the optimization observed at one target site could simply be a side effect of optimization at the other target site. From the differences in energy transport mechanism and the somewhat different site distribution of $\lambda_{m,3}^{ga}$ and $\lambda_{m,4}^{ga}$ (Table \ref{tab:reorg_ori_vs_ga}), it can be inferred that these two quantities are reasonably uncorrelated. This can be seen explicitly from Figure  \ref{fig:eta4_vs_eta3} where we have recast the data, showing the efficiencies are only weakly correlated.

\section{Concluding Remarks}
To summarize, we have proposed a plausible design principle for FMO which involves an effective interplay between the system and system-bath interaction. Energy transport mechanisms and pathways are combined with the fine-tuning of pigment-protein interaction in such a way that the (inevitable) site-varying system-bath interaction strengths are efficiently exploited to optimize energy transport. Interestingly, the optimization is observed not only for the commonly assigned target site of BChl 3, but for BChl 4 as well. More importantly, this optimization for the two target sites is largely uncorrelated, meaning that the observed optimization at BChl 4 is non-trivial.

In fact, it appears that the electronic system itself may be designed to transport energy to the two target sites via two different mechanisms of energy transport. For BChl 3, the optimal mechanism is purely dissipative funnelling while for BChl 4, the optimal mechanism is partially coherent. This implies that the system and system-bath interaction are configured in such a way that both dissipative and coherent processes are reasonably accommodated. Indeed, we observe an optimization of exciton delocalization when the FMO Hamiltonian is paired with the realistic configuration of site-dependent reorganization energies.

This simultaneous optimization of two largely uncorrelated processes of somewhat opposing nature is noteworthy since it suggests the presence of an underlying evolutionary design principle. It requires spatial ``engineering'' of system-bath interaction such that the magnitude of the reorganization energies either mutually benefits energy transport to both target sites or at least does not negatively impact energy transport to any of the target sites too significantly. Hence, if the simultaneous optimization at the two target sites is not purely accidental, and if coherence does indeed play a role in the FMO complex, then the conventional view of BChl 3 as the sole linker site to the reaction centre is possibly incomplete.

Finally, we would also like to comment briefly on the 180 \si{cm^{-1}} vibronic mode which has been shown to enhance electronic coherence in FMO \cite{chin2013role}. We note that this mode is also resonant with the energy difference between BChls 3 and 4. It is possible that this mode may be relevant to the design principle we have presented in this paper. We are exploring this possibility currently.

\begin{acknowledgments}
We thank Mi Kyung Lee for useful discussions and for providing the spectral density data used in this work. S. A. O. and D. A. W. H. gratefully acknowledge financial support from the Dodd-Walls Centre for Photonic and Quantum Technologies and the University of Otago. S. A. O. and D. A. W. H. also acknowledge the computational resources provided by the New Zealand eScience Infrastructure (NeSI).
\end{acknowledgments}

\appendix
\section{Modified Redfield Theory} \label{app:MRT}
The exciton population dynamics is described by the following rate equation:
\begin{eqnarray}
\dfrac{dP(t)}{dt} = K^{MRT}P(t),
\label{eq:modred}
\end{eqnarray}
where $K^{MRT}$ is the Modified Redfield rate matrix and $P(t) = [P_1(t) \ P_2(t) \dots P_k(t) \dots P_M(t)]^\intercal$ is a vector of exciton populations at time $t$, with the total number of exciton states $M = N$. The matrix element $K_{k,k'}^{MRT}$ of $K^{MRT}$ is the population transfer rate from the $k'$th to the $k$th exciton state and is given by the following time integral:
\begin{eqnarray}
K_{k,k'}^{MRT} = 2Re\int_0^{\infty} d \tau F_{k'}^*(\tau)A_k(\tau)N_{k,k'}(\tau),
\label{eq:R_diss}
\end{eqnarray}
\begin{equation*}
\begin{aligned}
\text{where} ~ F_{k'}(\tau) & = \text{exp} (-i(E_{k'}^0 - \lambda_{k'}) \tau - g_{k'k',k'k'}^*(\tau)), \\
A_k(\tau) & = \text{exp} (-i(E_{k}^0 + \lambda_{k}) \tau - g_{kk,kk}(\tau)), \\
N_{k,k'}(\tau) & = ( \ddot{g}_{k'k,kk'}(\tau) - [\dot{g}_{k'k,kk}(\tau) - \dot{g}_{k'k,k'k'}(\tau) \\
& - 2i\lambda_{k'k,k'k'}][\dot{g}_{kk',kk}(\tau) - \dot{g}_{kk',k'k'}(\tau) \\
& - 2i\lambda_{kk',k'k'}])e^{2(g_{kk,k'k'}(\tau) + i\lambda_{kk,k'k} \tau)}. \\
\end{aligned}
\end{equation*}

Here, $\lambda_{\delta \sigma , \kappa \mu}$ and $g_{\delta \sigma , \kappa \mu}(t)$ are the exciton counterparts of the site reorganization energy $\lambda_m$ and site lineshape function $g_m(t)$ respectively with the relationships expressed as:
\begin{eqnarray}
\lambda_{\delta \sigma , \kappa \mu} & = \sum_{m=1}^{N} a_{\delta \sigma}(m) a_{\kappa \mu}(m) \lambda_m ,
\end{eqnarray}
\begin{eqnarray}
g_{\delta \sigma , \kappa \mu}(t) & = \sum_{m=1}^{N} a_{\delta \sigma}(m) a_{\kappa \mu}(m) g_m(t),
\end{eqnarray}
where $a_{k,k'} = c^*_{m,k}c_{m,k'}$. $\lambda_k = \lambda_{kk,kk}$ is the $k$th exciton reorganization energy, which when subtracted from the $k$th exciton transition energy $E_k$ (i.e. the $k$th eigenenergy of $H_s$), gives the corresponding 0-0 transition energy, i.e. $E_{k}^0 = E_k - \lambda_k$. Meanwhile, the site lineshape function $g_m(t)$ is defined by the following double time integral:
\begin{eqnarray} \label{app:lineshape_function}
g_m(t) = \int_0^t dt_1 \int_0^{t_1} dt_2 C_m(t_2),
\end{eqnarray}
where $C_m(t)$ is the complex bath correlation function which can be written as:
\begin{multline} \label{eq:bath_corr_function}
C_m(t) = \dfrac{1}{\pi} \int_0^{\infty} d\omega J_m(\omega)\left[ \text{cos}(\omega t) \text{coth}\left(\dfrac{\beta \omega}{2}\right) \right.\\
	                               			 				 \left. - \ i \ \text{sin}(\omega t) \vphantom{\text{cos}(\omega t) \text{coth}\left(\dfrac{\beta \omega}{2}\right)}\right].
\end{multline}
Thus, $g_m(t)$ can be computed from the spectral density $J_m(\omega)$ from the following relation:
\begin{multline}
g_m(t) = \dfrac{1}{\pi} \int_0^{\infty} d\omega \dfrac{J_m(\omega)}{\omega^2} \left[ (1 - \text{cos} \ \omega t) \text{coth}\left(\dfrac{\beta \omega}{2}\right) \right. \\
\label{eq:lineshape_function}																						\left.	+ \ i(\text{sin} \ \omega t - \omega t) \vphantom{\text{coth}\left(\dfrac{\beta \omega}{2}\right)} \right].
\end{multline}
The integrand in eq. \ref{eq:lineshape_function} contains poles, and thus the lower limit of integration must be changed to an appropriately small finite value for numerical integration. Alternatively, the poles can be removed at the level of $C(t)$ by analytical means, such as in the case of the Drude-Lorentz representation. The derivation of the analytical form of $g_m(t)$ for the Drude-Lorentz spectral density is shown in Appendix \ref{app:drude_lineshape}.

Lastly, the diagonal elements of $K^{MRT}$ are computed from:
\begin{eqnarray}
K_{k,k}^{MRT} = -\sum\limits_{k'} K_{k',k}^{MRT}.
\end{eqnarray}

Eq. \ref{eq:modred} can then be easily solved with $P(t) = e^{K^{MRT}t}P(0)$. Here the initial population vector consists of the exciton populations corresponding to the initial excitation site $m_0$, i.e. $P(0) = [c_{m_0,1}^2 \ c_{m_0,2}^2 \dots \ c_{m_0,k}^2 \dots c_{m_0,M}^2]^\intercal$.

It should be pointed out that the numerical integration in eq. \ref{eq:R_diss} requires a finite cutoff value for the upper integration limit. In general, the stronger the system-bath coupling, the more oscillatory the integrand and the longer it takes for the integrand function to taper off. We have taken care to ensure our chosen cutoff value of $\tau = 1 \ \si{ps}$ is acceptable by checking that increasing the cutoff has negligible effect on the calculated rates. For the genetic algorithm and statistical evaluation performed in this work, such manual inspection would be impractical. As such, we have applied a ``worst case scenario" test by assigning $\lambda_m$ with the largest reorganization energy to each site and confirmed there is no appreciable change in the computed rates when the value of the cutoff is increased to 5 \si{ps^{-1}}.

\section{Coherent Modified Redfield Theory}   \label{app:CMRT}
The formula for the exciton transfer rates for CMRT is similar to eq. \ref{eq:R_diss}, except with the upper limit of integration replaced by $t$ to incorporate non-Markovianity. That is, the rates are time-dependent in contrast to the Markovian MRT case. We shall denote this exciton population transfer rate (from exciton $k'$ to $k$) as $K_{k,k'}^{CMRT}(t)$. Due to the inclusion of coherences in CMRT, dephasing rates must also be accounted for and is defined by the following relation:

\begin{eqnarray}
L_{kk'}(t) &= \sum\limits_{m}\Big[ a_{kk}(m) - a_{k'k'}(m) \Big]^2 Re \Big[ \dot{g_m}(t) \Big].
\label{eq:R_deph}
\end{eqnarray}

It turns out the CMRT master equation can be cast into a generalized Lindblad form:
\begin{align}
\dfrac{d \rho(t)}{dt} = &-i\Big[\hat{H_e},\rho(t) \Big] - \dfrac{1}{2}\sum_{k,k'} R_{kk'}(t) \Big[ \Big\{ A_{kk'}^{\dagger}A_{kk'},\rho(t) \Big\} \notag \\
& - 2A_{kk'} \rho(t) A_{kk'}^{\dagger} \Big].
\label{eq:cmrt_lindblad}
\end{align}

This allows the master equation to be conveniently solved using the Non-Markovian Quantum Jump (NMQJ) approach. The summary of the NMQJ technique is provided in Appendix \ref{app:nmqj} while full details and derivation can be found in Refs. \onlinecite{piilo2008non,piilo2009open,breuer2009stochastic}. In eq. \ref{eq:cmrt_lindblad}, $A_{kk'} = \vert k \rangle \langle k' \vert $ are the jump operators while $H_e = \sum_k E_k^0 \ket{k} \bra{k}$ is the modified system Hamiltonian that governs the coherent evolution. The jump rates $R_{kk'}(t)$ from exciton state $k'$ to $k$ are such that: \\
\begin{eqnarray}
R_{kk'}(t) =
\begin{cases}
  G_k(t)     &\text{for } k = k',\\
  K_{kk'}^{CMRT}(t)  &\text{for } k \neq k',\\
\end{cases}
\label{eq:R_kk'_matrix_elements}
\end{eqnarray}
where $G_k(t)$ is the $k$th element of vector $G(t)$ which is linked to the pure dephasing rate $L_{kk'}(t)$. This relation is given by $G(t) = B^{-1}D(t)$ where $B$ is a matrix and $D(t)$ is a time-dependent vector whose elements are defined respectively as:
\begin{eqnarray}
B_{kk'} =
\begin{cases}
  0.5     &\text{for } k' < k,\\
  0.5(2M - k')  &\text{for } k = k',\\
  1     				&\text{for } k < k' < M,\\
  0.5 &\text{otherwise},
\end{cases}
\end{eqnarray}
\begin{eqnarray}
D_a(t) = \sum_{k=a+1}^{M}L_{ak}(t) + \sum_{k=1}^{M-1}L_{ka}(t).
\end{eqnarray}

\section{Non-Markovian Quantum Jump} \label{app:nmqj}

The Non-Markovian Quantum Jump (NMQJ) is a tool for the stochastic unravelling of a non-Markovian quantum master equation, and is the non-Markovian generalization of the well-known Monte Carlo Wave Function (MCWF) method \cite{dalibard1992wave,molmer1993monte}. It takes advantage of the general definition of the density matrix,
\begin{eqnarray} \label{eq_app:density_matrix_define}
\rho(t) = \sum\limits_{\alpha} \dfrac{N_{\alpha}(t)}{N}\ket{\psi_{\alpha}(t)}\bra{\psi_{\alpha}(t)},
\end{eqnarray}
where $N$ is the ensemble size and $N_{\alpha}$ is the number of ensemble members in the state $\ket{\psi_{\alpha}(t)}$, and operates on the level of the state vector.

In this work, the $\ket{\psi_{\alpha}(t)}$s are the exciton states $\ket{k}$ and the time-evolved initial state $\ket{\psi_0(t)}$. $\ket{\psi_0(t)}$ is a coherent superposition of exciton states in which $\ket{\psi_0(0)} = \ket{m_0}$, where $\ket{m_0}$ is the initial localized site basis. Due to the construct of the density matrix, the single exciton states are time-independent as global phases are cancelled out. Hence eq. \ref{eq_app:density_matrix_define} can be rewritten as:
\begin{multline} \label{eq_app:density_matrix_define2}
\rho(t) = \dfrac{N_0(t)}{N}\ket{\psi_0(t)}\bra{\psi_0(t)} + \sum\limits_{k}\dfrac{N_{k}(t)}{N}\ket{k}\bra{k}.
\end{multline}
Note that $\rho(t)$ is in the site basis due to eq. \ref{eq:M_superposition_of_m}.

According to the NMQJ formulation, each ensemble member undergoes continuous time evolution interrupted by discontinuous probabilistic jumps. The propagation of the state vector proceeds in small time steps $\delta t$ with the deterministic evolution described by
\begin{eqnarray}
\ket{\psi_{\alpha}(t + \delta t)} = \dfrac{e^{-i\hat{H_{eff}}\delta t}\ket{\psi_{\alpha}(t)}}{\left\|e^{-i\hat{H_{eff}}\delta t}\ket{\psi_{\alpha}(t)}\right\|},
\end{eqnarray}
where the effective non-Hermitian Hamiltonian is defined as
\begin{eqnarray}
\hat{H_{eff}}(t) = \hat{H_e} - \dfrac{i}{2}\sum\limits_{k,k'} R_{kk'}(t)A_{kk'}^{\dagger}A_{kk'}.
\end{eqnarray}

When $R_{kk'}(t) \geq 0$, an instantaneous positive jump to another state $\ket{\psi_{\alpha'}(t)}$ may occur, and the state at the next time step would be set to this new state.:
\begin{eqnarray}
\ket{\psi_{\alpha}(t)} \rightarrow \dfrac{A_{kk'}\ket{\psi_{\alpha}(t)}}{\left \| A_{kk'}\ket{\psi_{\alpha}(t)} \right \|} = \ket{\psi_{\alpha'}(t + \delta t)}.
\end{eqnarray}
Here the probability of the jump occurring through the channel $k' \rightarrow\ k$ for a given ensemble member state $\ket{\psi_{\alpha}}$ is
\begin{eqnarray}
P_{\alpha,kk'}^+(t) = R_{kk'}(t) \delta t \bra{\psi_{\alpha}(t)} A_{kk'}^{\dagger}A_{kk'}\ket{\psi_{\alpha}(t)}.
\end{eqnarray}

Unlike purely Markovian dynamics, the non-Markovian transition rate $R_{kk'}(t)$ is time-dependent and can become negative. During this time period, the action of the positive jump operator $A_{kk'}$ is to bring the target state $\ket{\psi_{\alpha'}(t)}$ to the source state $\ket{\psi_{\alpha}(t)}$:
\begin{eqnarray}
\ket{\psi_{\alpha'}(t+ \delta t)} \leftarrow \ket{\psi_{\alpha}(t)} = \dfrac{A_{kk'}\ket{\psi_{\alpha'}(t)}}{\left \| A_{kk'}\ket{\psi_{\alpha'}(t)} \right \|}.
\end{eqnarray}
This implies that the negative jump operator $A_{kk'}^{-} = \ket{\psi_{\alpha'}(t)}\bra{\psi_{\alpha}(t)}$. In other words, a negative jump means the reversal of a previous jump back to a prior state. The probability for a reverse jump is given by:
\begin{multline}
P_{\alpha\rightarrow\alpha',kk'}^-(t) = \dfrac{N_{\alpha '}(t)}{N_{\alpha}(t)}\vert R_{kk'}(t)\vert \delta t \\
\times \bra{\psi_{\alpha '}(t)} A_{kk'}^{\dagger}A_{kk'}\ket{\psi_{\alpha '}(t)}.
\end{multline}
With each jump, the number of ensemble members in the source and target states are updated accordingly for the current time step, i.e. $N_{\alpha}(t) - 1$ and $N_{\alpha'}(t) + 1$ respectively.

The choice between deterministic evolution or jump is determined by a random number $0 < \epsilon < 1$. If $\epsilon$ is less than or equals to the total jump probabilities of all channels, a jump occurs and vice versa. If a jump is determined, another random number $s$ is generated to randomly select the jump channel.

At $t = 0$, all the ensemble members are in $\ket{\psi_0(0)}$, i.e. $N_0(0) = N$. This means the propagation of the density matrix starts from a pure state and progresses to a mixed state through positive jumps. At a later time when the rates become negative, negative jumps can undo these positive jumps, and this may include the revival of coherences via a reverse jump to $\ket{\psi_0(t)}$. For FMO, there are 56 relaxation and 8 dephasing channels. As can be seen from eq. \ref{eq_app:density_matrix_define2}, efficient computation and averaging of the constituent density matrices can be achieved by simply updating $N_{\alpha}(t)$ and $N_{\alpha'}(t)$ at each time step and performing a one-off time evolution of $\ket{\psi_0(t)}$.

\section{Analytical form of the Drude-Lorentz lineshape function} \label{app:drude_lineshape}
Since the Drude-Lorentz spectral density $J_m^{DL}(w)$ has simple poles at $\omega = \pm i \Omega_m$, the residue theorem can be conveniently applied to obtain the analytical form of the bath correlation function $C_m(t)$ \cite{kleinekathofer2004non} which is:
\begin{eqnarray}
C_m(t) = \sum_{k=1}^{n_r} \alpha_k^r e^{\gamma_k^r t} - i \alpha^i e^{\gamma^i t},
\end{eqnarray}
where
\begin{eqnarray*}
\alpha_k^r =
\begin{cases}
\lambda_m\Omega_m \text{cot}\left(\dfrac{\beta \Omega_m}{2}\right)    &\text{for } k = 1, \\ \\
-\dfrac{4 \lambda_m}{\beta \Omega_m} \ \dfrac{\nu_{k-1}}{1 - \left(\nu_{k-1}/\Omega_m\right)^2}  &\text{for } k = 2 \ \text{to } n_r,
\end{cases}
\end{eqnarray*}

\begin{eqnarray*}
\gamma_k^r =
\begin{cases}
-\Omega_m    &\text{for } k = 1, \\ \\
-\nu_{k-1}  &\text{for } k = 2 \ \text{to } n_r,
\end{cases}
\end{eqnarray*}

\begin{eqnarray*}
\alpha^i = \lambda_m \Omega_m, \\
\gamma^i = -\Omega_m.
\end{eqnarray*}

Here $\nu_k = \dfrac{2\pi k}{\beta}$ are the Matsubara frequencies. In principle, the Matsubara expansion is infinite but in practice, the summation can be truncated at some finite value $n_r$. The number of terms required for convergence is dependent only on temperature, with more terms needed for lower temperatures.

Substituting into eq. \ref{app:lineshape_function}, we arrive at the analytical form of the lineshape function:

\begin{multline}
g_m(t) = \sum_{k=1}^{n_r} \dfrac{\alpha_k^r}{(\gamma_k^r)^2}(e^{\gamma_k^r t} - \gamma_k^r t - 1) \\
- i\dfrac{\alpha^i}{(\gamma^i)^2}(e^{\gamma^i t} - \gamma^i t - 1).
\end{multline}
\clearpage

\section{Histogram of efficiencies for random configurations of $\lambda_m$} \label{app:hist_rand_complete}

\begin{figure}[htp]
    \subfloat[]{\includegraphics[scale=0.7]{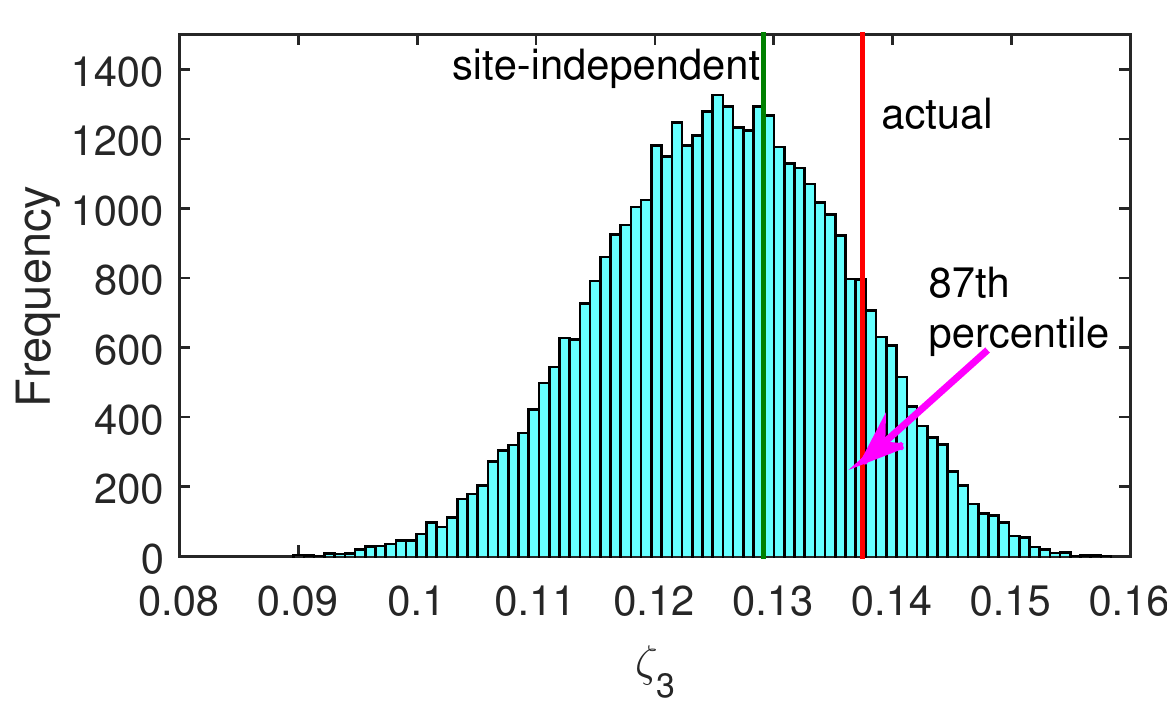}} \\
    \subfloat[]{\includegraphics[scale=0.7]{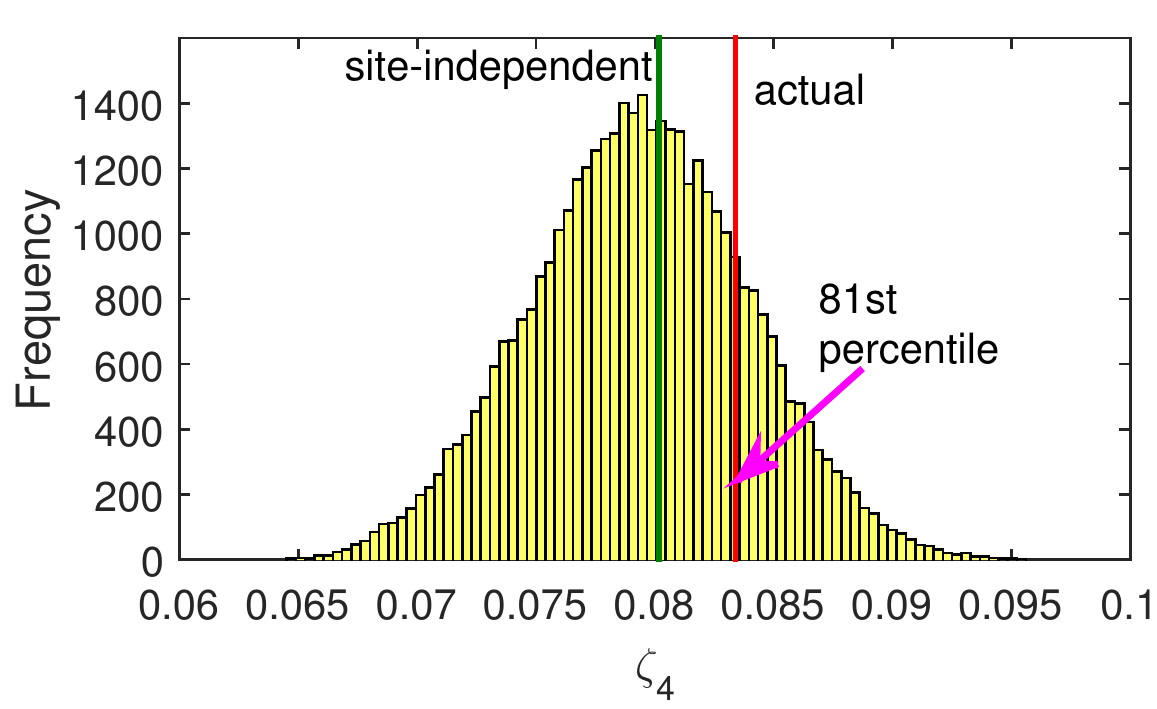}}
    \caption{\label{fig:hist_rand_complete}
             Histogram of efficiencies $\zeta_{m}$ for random configurations of
             $\lambda_m$ with the same mean, maximum and minimum as the original
             for exciton state (corresponding target site) of (a) $k=1$ (BChl 3)
             and (b) $k=2$ (BChl 4). In both histograms, the original
             configuration is also included in the sample and the sample size is
             the same as in Fig.~\ref{fig:hist_rand_perm}, i.e. 40320. The red
             and green vertical lines indicate the position on the histogram of
             the original and site-independent configurations respectively.
             Numerical method: MRT.}

\end{figure}

\section{Numerical details of the genetic algorithm} \label{app:gen_algo}

We have utilized the Genetic Algorithm solver from the MATLAB optimization
toolbox with the following parameters: (i) Population size: 80, (ii) Selection
function: Stochastic uniform, (iii) Elite count: 4, (iv) Crossover function:
Intermediate, (v) Crossover fraction: 0.8, (vi) Mutation function = Adaptive
feasible, (vii) Number of generations: 4000. To maximize the efficiency of the
algorithm while representing a realistic scenario at the same time, we confined
the search space to be within the minimum and maximum values of the original
$\lambda_m$. The choice of the upper bound is also important for another reason:
in order for a positive correlation between larger $\lambda_m$ and higher
efficiency to stay valid for a dissipative process, the allowed values of
$\lambda_m$ must not be so large to the point that the quantum Zeno effect
becomes relevant \cite{mohseni2008environment}.

\bibliography{fmo_paper}

\end{document}